\documentclass[dvips]{article}
\usepackage{icrctc07,amsmath,amssymb,epsfig,color}

\title{Antideuterons from supersymmetric dark matter}
\shorttitle{Antideuterons from supersymmetric dark matter}

\authors{Fiorenza Donato$^{1}$, Nicolao Fornengo$^{}$, David Maurin$^{3}$ }
\shortauthors{Antideuterons from supersymmetric dark matter}
\afiliations{$^1$Dipartimento di Fisica Teorica, Universit\'a di Torino and INFN, 
via P. Giuria 1, 10125 Torino, Italy \\ 
$^2$Laboratoire de Physique Nucl\'eaire et Hautes Energies, CNRS-IN2P3/Universit\'es
Paris VI et Paris VII, 4 place Jussieu, Tour 33, 75252 Paris Cedex 05\\ }
\email{donato@to.infn.it, fornengo@to.infn.it, dmaurin@lpnhe.in2p3.fr}

\abstract{We calculate the antideuteron flux expected from dark matter annihilation
in the galactic halo. The propagation is treated in a full 2-D propagation model
consistent with the results obtained from the propagation of B/C and other galactic
species. We discuss the potentials of this indirect dark matter detection means, with special
emphasis on the possible sources of uncertainties affecting future measurements.}

\begin{document}
\maketitle

\section{Introduction}

Antideuterons have been proposed as an interesting indirect detection signal of particle dark matter (DM) in 
Ref. \cite{2000PhRvD..62d3003D}, where it has been shown that the low energy part of the antideuteron spectrum
is potentially able to offer a good signal--to--noise gain. In this paper we present a novel evaluation of the antideuteron flux, with special emphasis on the determination of the astrophysical uncertainties arising from
antideuteron propagation in the galactic medium. To this aim, we take advantage of a more refined and constrained diffusion model \cite{2001ApJ...555..585M}, as well as updated cross sections for the calculation of the antideuteron background \cite{2005PhRvD..71h3013D}. Our approach follows closely that of Ref. \cite{2004PhRvD..69f3501D}, which focussed on the antiproton component in cosmic-rays as a dark matter signal.


\section{Dark matter source}

We model the DM distribution in the following general form:
\begin{equation}
\rho= \rho_s  \left( \frac{r}{r_s}\right)^{-\gamma}\left[
	  1+\left(\frac{r}{r_s}\right)^{\alpha}\right]^{(\gamma-\beta)/\alpha},
\label{profile}
\end{equation}
where $\rho_s$ and $r_s$ denote the scale density and scale radius.
The mean slope of the cusp obtained from numerical simulations is well fitted
by a $(\alpha=1,\beta=3,\gamma)$ profile, with $\gamma= 1.16\pm0.14$ \cite{2004MNRAS.353..624D}.
In this paper, we set $\gamma=1.2$. The exact slope of the density profile
is not crucial regarding the calculation of antiproton\ and antideuteron\ DM-induced
fluxes~\cite{2003A&A...404..949M,2004PhRvD..69f3501D}, as well as a possible spike in the central region (that is certainly crucial in the context of $\gamma$-ray indirect detection). The possible
effect of clumpiness is not addressed here \cite{lavalle}.
The parameters of the Milky Way dark matter distribution are set
to $r_s=20$~kpc and $\rho_0= 0.3$~GeV~cm$^{-3}$, according to typical
values found in kinematical data in the Galaxy \cite{2006MNRAS.370.1055B}
or for simulated haloes \cite{2007ApJ...657..262D}.

The production of antideuterons from the annihilation of a neutralino 
pair is based roughly on two steps: the hadronization into a antiproton and 
antineutron pair, followed by their subsequent fusion into a antideuteron bound system.
The hadronization has been computed as in Ref. \cite{2004PhRvD..69f3501D}, 
while for the nuclear fusion we adopt the simple coalescence scheme 
as described in Ref. \cite{2000PhRvD..62d3003D} and \cite{2005PhRvD..71h3013D}.
We fix the coalescence parameter to be $P_{\rm coal}$= 79 MeV both for primary
and secondary fluxes.

\section{Propagation Model}

Charged particles travelling from
the sources to the solar neighborhood are affected by
the scattering off random magnetic fields 
(diffusion in energy and space) and the Galactic
wind (convection).
We follow closely the description given in Ref. \cite{2001ApJ...555..585M}.
The framework is a diffusion--convection--reacceleration model,
with a spatial independent diffusion coefficient 
$K(E)=\beta K_0{\cal R}^{\delta}$
(where ${\cal R}=pc/Ze$ is the rigidity) and a constant wind $V_c$ directed
outwards along $z$. Cosmic rays are confined within a diffusive halo of helf--height $L$
and within the radius $R=20$~kpc of the Galaxy.
Energy losses (Coulomb, ionization, adiabatic) and gains
(reacceleration) are included in a disk of thickness $2h=200$~pc.
The antideuteron spectra are calculated at the Earth position in the galactic disk, which we fix at 
$R_\odot=8.0$~kpc.

The five key parameters of the model are the halo size $L$ of the Galaxy, 
the normalization of the diffusion coefficient $K_0$ and its slope
$\delta$, the constant galactic wind $V_c$ and the level of reacceleration
through the Alfv\'enic speed $V_a$.
These parameters, constrained by the analysis of the B/C ratio, reflect degenaracies in the parameters space.
\cite{2001ApJ...555..585M,2002A&A...394.1039M}.
For antiprotons, all sets of parameters {\em consistent with B/C} lead to the very same secondary flux
\cite{2001ApJ...563..172D}, but to a large scatter for the primary (from DM sources) fluxes
\cite{2004PhRvD..69f3501D}. Unsurprisingly, the same pattern is observed for antideuterons.
Table~\ref{table:prop} gathers the parameters used to calculate
the standard antideuteron\ flux (med), as well as the two sets of parameters leading
respectively to the minimal (min) and maximal (max) exotic fluxes \cite{2004PhRvD..69f3501D}.
\begin{table}[t]
\begin{center}
{\begin{tabular}{ccccc}
\hline
{\rm case} &  $\delta$  & $K_0$                 & $L$   & $V_c$    \\
           &            & [${\rm kpc^{2}/Myr}$] & [kpc] & [km/sec] \\
\hline
\hline
{\rm max} &  0.46  & 0.0765 & 15 & 5    \\
{\rm med} &  0.70  & 0.0112 & 4  & 12   \\
{\rm min} &  0.85  & 0.0016 & 1  & 13.5 \\
\hline
\end{tabular}}
\caption{
Relevant sets of propagation parameters used in the analysis (the value of $V_a$, unimportant
for the primary calculation, is omitted). 
\label{table:prop}}
\end{center}
\end{table}
%

\section{Cross sections for the background flux}
\label{sec:Xsec}

\begin{figure}[t]
\begin{center}
\noindent
\includegraphics [width=.5\textwidth,clip]{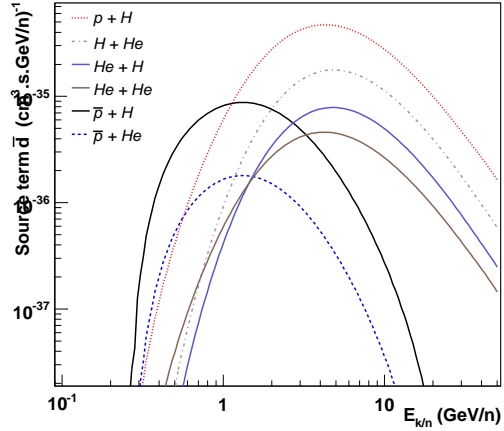}
\end{center}
\caption{Antideuteron source term 
for the secondary flux, obtained from the integration of the measured CR fluxes times
the differential production cross section, times the gas density in the ISM. The contributions
of the different processes are outlined, as well as the total flux.}
\label{fig:separated_contrib}
\end{figure}

\begin{figure}[t]
\begin{center}
\noindent
\includegraphics [width=.5\textwidth,clip]{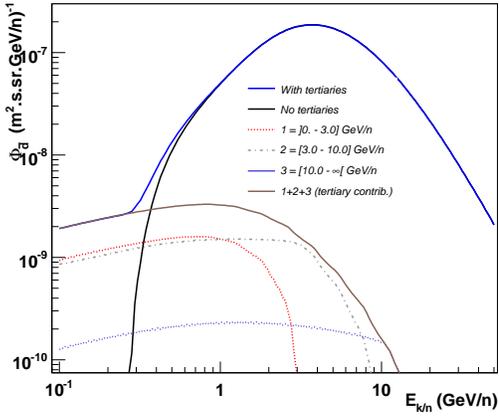}
\end{center}
\caption{Interstellar antideuteron\ flux with and without the tertiary source term.
The top curve is the sum of the secondary contribution (steep
decrease at low energy) and of the tertiary contribution (a few
secondaries survive the collision with the ISM and are shifted to
lower energy). The tertiary curve, labeled ``1+2+3", is the sum of the three curves
``1", ``2" and ``3" that correspond to the tertiary 
contributions of three energy bands (labelled as: 1, 2 and 3).}
\label{fig:tertiary_flux}
\end{figure}

The cross sections for nuclear interactions are taken from Ref. \cite{2005PhRvD..71h3013D}.
The contributions of the various channels are displayed in Fig.~\ref{fig:separated_contrib}.
The $pH$ and $p He$ are the dominant channels for energies beyond 1 GeV/n.
At lower energies, the $\bar{p}p$ contribution plays an
important role, as it enhances the background,
in a region where the primary contribution is expected.
All cross sections can be assumed to be
accurate  within a factor of two,
except for the $\bar{p}p$ and $\bar{p} He$ channels
which  are not yet well constrained.

The tertiary contribution (due to non--annihilating rescattering
of antideuterons) is also implemented. As in Ref. \cite{2005PhRvD..71h3013D},
we use here a more refined treatment  than the classical Tan \& Ng scheme (energy of the surviving
antinuclei scales as the energy of the incoming antideuteron). Fig. \ref{fig:tertiary_flux} shows that
tertiaries are relevant only at very lowe energies, below 0.3 GeV/n. Taking into account energy redistributions and the solar modulation of the flux washes out the effect of tertiaries almost completely for detectable fluxes at the top of the Earth atmosphere. This is manifest in Fig. \ref{fig:comp_data}, where the background calculation is shown after solar modulation at the minimun of solar activity has been included.

We wish to comment that we have corrected here an error in the normalization of the cross
sections used in previous calculations (Fig.~8 of Ref. \cite{2005PhRvD..71h3013D} is incorrect).
However this modification does not alter the conclusions of Ref. \cite{2005PhRvD..71h3013D}.


\section{Signal from dark matter annihilation}
\label{sec:results}

\begin{figure}[t]
\begin{center}
\noindent
\includegraphics [width=.5\textwidth,clip]{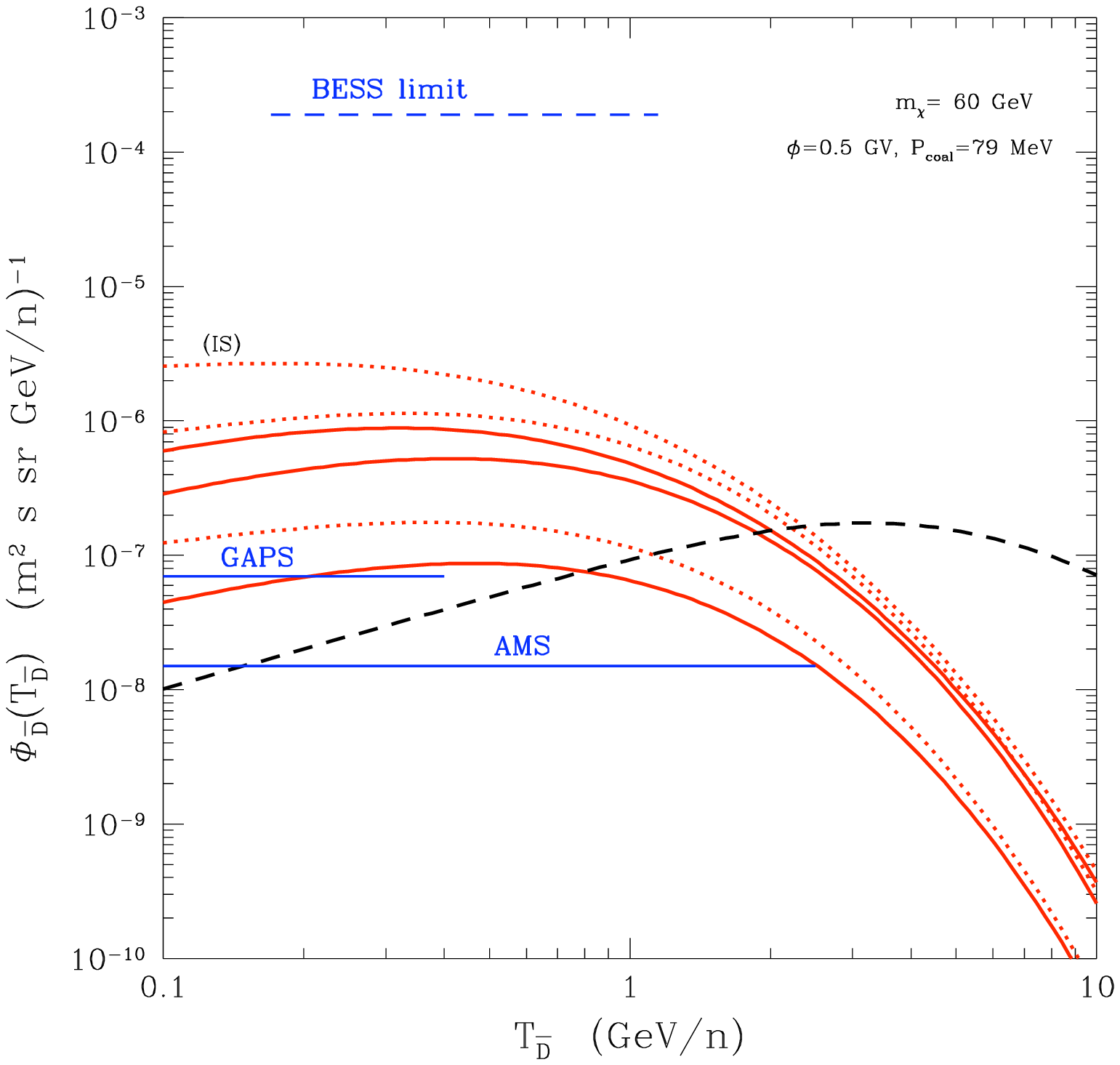}
\end{center}
\caption{Fluxes of primary and secondary antideuterons at solar minimum.
The dashed line shows the secondary 
antideuteron contribution, calculated with the median set of propagation parameters. 
The solid lines correspond to the fluxes from the annihilation of a 60 GeV DM particle.
From top to bottom: the maximum, median and minimum 
set of propagation parameters are used. 
Dotted lines: the same as for solid lines, but for the interstellar fluxes.
The BESS limit on the antideuteron searches is shown as a horizonthal dashed line. The
expected sensitivities for the GAPS and AMS experiments are shown as horizonthal solid lines.}
\label{fig:comp_data}
\end{figure}

\begin{figure}[t]
\begin{center}
\noindent
\includegraphics [width=.5\textwidth,clip]{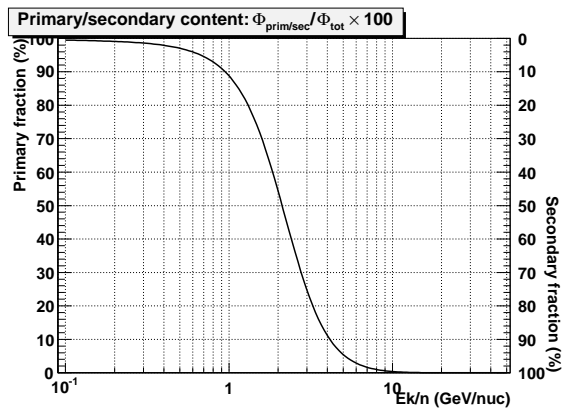}
\end{center}
\caption{Ratio of the primary--to--secondary fluxes, calculated for the set
of maximal propagation parameters and at solar minimum. The DM primary flux 
has been obtained for a particle with a mass of 60 GeV.} 
\label{fig:primarycontent}
\end{figure}

An example of antideuteron signal from DM annihilation is shown in Fig. \ref{fig:comp_data}, for a 60 GeV--mass DM particle with an annihilation cross section compatible with the current WMAP result on the amount of cold DM. The galactic propagation has been calculated for the three sets of astrophysical parameters shown in Table 1. We see that, as for antiprotons \cite{2004PhRvD..69f3501D} and for positrons \cite{positrons}, the uncertainty in the low--energy part of the signal, as due to galactic propagation, is sizeable. It reaches one order of magnitude at energies of 0.1 GeV/n. Nevertheless, we find, as already discussed in Ref. \cite{2000PhRvD..62d3003D}, that the low--energy sector of the antideuteron signal potentially offers a good discriminating power over the backgound, a feature wich is not shared neither by the antiproton \cite{2004PhRvD..69f3501D} nor the positron \cite{positrons} signals. For this reason, antideuterons represent the best indirect detection option, with a potential clear signature in the low energy part of the spectrum. An example of the signal--to--noise ratio is show in Fig. \ref{fig:primarycontent} for the same representative signal of Fig. \ref{fig:comp_data}: we see that below 1-2 GeV/n energies, the total antideuteron flux at the Earth is dominated by the signal component.
Clearly the optimistic situation shown in Figs.\ref{fig:comp_data} and \ref{fig:primarycontent} depends also on the particle physics properties of the DM candidate, mostly on its abundance as DM component. Nevertheless, antideuterons offer the possibility for detectors like GAPS and AMS (whose expected sensitivities are reproduced in Fig. \ref{fig:comp_data}) to explore a large fraction of the particle--physics parameter space in case of  neutralino DM in supersymmetric theories. These results will be discussed elsewhere \cite{newdbar}.


\section{Conclusions}

We have discussed a novel calculation of the antideuteron component of cosmic rays, by introducing additional effects in the background estimate and by discussing the astrophysical uncertainties coming mainly from the antideuteron propagation in the galactic medium, which affect mostly the DM signal contribution. As for antiprotons, the primary component from DM annihilation is more affected than the secondary component from cosmic rays interactions with the interstellar medium. The uncertainty on the signal is about one order of magnitude in the low energy part of the spectrum. Prospects of future forthcoming experiments have been discussed: for suitable DM--particle properties, the antideuteron flux may well fall within the experimental reach, regardless of the astrophysical uncertainties. 

\section{Acknowledgements}
N.F. and F.D. aknowledge research grants funded jointly by the italian Ministero dell'Universit\`a e della Ricerca (MIUR), the University of Torino and the INFN within the {\sl Astroparticle Physics Project}.


\bibliography{icrc0519}
\bibliographystyle{plain}

\end{document}